# AB 5 type multicomponent TiVCoNiMn2 high-entropy alloy

**Abhishek Kumar[1], M. A. Shaz[1], N.K. Mukhopadhyay[3] and Thakur Prasad Yadav[1, 2*]**

[1]Hydrogen Energy Centre, Department of Physics, Institute of Science
Banaras Hindu University, Varanasi-221005, India

[2]Department of Physics, Faculty of Science, University of Allahabad, Prayagraj-211002, India

[3]Department of Metallurgical Engineering, Indian Institute of Technology (Banaras Hindu University), Varanasi-221005, India

*Email: tpyadav@allduniv.ac.in

**Abstract**

Recent theoretical and practical research has focused on multi-component High Entropy Alloys (HEAs), which have superior mechanical and functional properties than standard alloys based on a single major element, thereby establishing a new field. A multi-component HEA contains five or more primary elements at concentrations ranging from 5 to 35 atomic percent. We examined the microstructure and mechanical properties of TiVCoNiMn$_2$ HEA. The mixing enthalpy and other thermodynamic parameters were determined using Meidma's model. TiVCoNiMn$_2$ exhibits a mixing enthalpy of -15.6 kJ/mol and an atomic radius mismatch of approximately 10.03%. This HEA was produced using a 35-kW radio frequency induction furnace. XRD profile analysis confirms the BCC phase with a lattice parameter of 2.95Å. Additional intermetallic phases are evaluated after 24 hours of annealing at 1000°C. SEM and EDX were used to analyze the elemental composition and surface morphology of the as-cast and annealed HEAs. The hardness was tested at 200 g with a dwell time of 15 minutes. Hardness did not differ between annealed (795 HV) and as-cast (806 HV). HEA is derived from both hydride and non-hydride-producing elements. This could be a useful hydrogen storage material. The hydrogen absorption/desorption capabilities of these HEAs are promising.

## 1. Introduction:

Multicomponent high entropy materials (HEMs), a unique material group, have recently garnered attention [1-4]. Unlike conventional alloys, multicomponent high entropy alloys (HEAs) have five or more elements with 5 to 35 atomic percent (at. %) concentrations. These alloys have high mixing entropy, stabilizing solid solution [5, 7]. Due to the high entropy effect, high-entropy alloys (HEAs) have a simple microstructure contrary to expectations. This effect simplifies microstructure and accelerates solid solution formation. Through defect engineering, alloy mechanical properties can be improved by adding elements with different atomic radii. Inter-elemental reactions and lattice deformation strengthen HEAs. HEAs have slower phase transition and diffusion rates than conventional alloys. Two main factors cause lattice distortion in HEAs. In HEA synthesis, different-sized elements are combined to create a lattice with different elemental sizes. This size difference causes larger atoms to press or displace smaller ones. Second, lattice distortion increases strain energy and HEA-free energy. The phenomenon affects the physical properties of HEAs, including solid solution strengthening. The high mixing entropy, which can reduce the number of phases in high-order alloys and improve their properties [8, 9]. High-entropy mixing is favorable for stabilizes disordered solid solution phases with simple crystal structures like BCC, FCC, and HCP [10].

The hysteretic diffusion effect stabilizes stable HEAs microstructures. Most FCC HEAs are soft but tough [11]. BCC-type HEAs are hard and brittle [12, 13]. Structure is major in HEAs formability, hardness, and strength. The leading role for formatin the phase is thermodynamics, atomic size mismatch, Allen electronegativity, mixing entropy, and mixing enthalpy affect the structure of HEAs [14, 15]. High pressure can cause phase transformation in HEAs [16]. This study examined how aluinum (Al) affects the high-entropy alloy Al$_x$CoCrFeNi. Their research revealed a major structural change. Germanium (Ge) enhancement in (CoCrFeNi)$_{100-x}$Ge$_x$ HEA was studied in a case study [17]. The study found a phase transformation from face-centered cubic (FCC) to body-centered cubic (BCC) structures in the 10≤x≤25 compositional range. In the Report study of structural and phase formation study in TiVZrCrNi, they get that annealing of as cast HEA did not form any other intermetallic phase despite the C14 Laves phase, but their mechanical properties enhanced [2].

In this present study is focused on synthesis, microstructure, and hardness evolution in BCC phase-based AB$_5$ type TiVCoNiMn$_2$ High Entropy Alloy (HEA). The mixing enthalpy and atomic radius mismatch were determined using Meidma's model, and we found that these values are favorable for BCC phase formation in this HEA and the TiVCoNiMn$_2$ exhibits a mixing enthalpy of -15.6 kJ/mol and an atomic radius mismatch of approximately 10.03%. This study is to choose the elements that can form hydrides and investigate the formation and stability of the BCC phase or other potential phases in the as-cast, annealed AB$_5$ type TiVCoNiMn$_2$ high-entropy alloy (HEA). We have successfully shown the formation of a single BCC phase that includes all five elements. No other significant phases were detected in the as-cast high-entropy alloy after annealing at 1000 °C for 24 hours, the evolution of some other intermetallic phase. We get the evolution of the intermetallic phase.

## 2. Experimental Section:
### 2.1 Material synthesis and characterization methods:

Alfa Aesar supplied high-purity powder for TiVCoNiMn$_2$ High Entropy Alloy (HEA) synthesis. The materials had a purity level of approximately ~99.50%. To synthesize a palette, elements were chosen by stoichiometry. The palette was made in a cylindrical steel mold with a hydraulic press that could apply $3 \times 10^5$ N/m$^2$. A palette weighing 10 g was used to synthesize a multi-component HEA using RF induction melting. The synthesis occurred in a ~99.90% pure argon atmosphere. The schematic digram shon in figure 1. Remelting of HEAs of four times ensures chemical homogeneity in ingots. As-cast induction melted HEA ingots were crushed and powdered for characterization. Phase analysis begins with Malvern Panalytical's Empyrean XRD system. This cutting-edge system has a 256x256-pixel area detector. The system uses a Cu radiation source (CuKa; = 1.5406) and a graphite monochromator. BRAGG-BRENTANO GEOMETRY uses 45 kV and 40 mA for XRD. This study used the TECNAI 20 G2 TEM at 200 kV to acquire sample microstructures and SAED patterns. As-prepared samples were examined for surface morphology using the EVO 18 SEM. Under a 10$^{-5}$ torr vacuum, the SEM operated at 25 kV. Another cold-mounted portion was hardened. Grinding machines polished the mounted ingot. The mounted HEAs' smooth surface was cloth-polished last. Due to its hardness, we used water and alumina powder to polish our cloth to a bright surface. After polishing, VIDAS Truement was used to test Vickers hardness with a 15-second dwell time and changing indenter load.

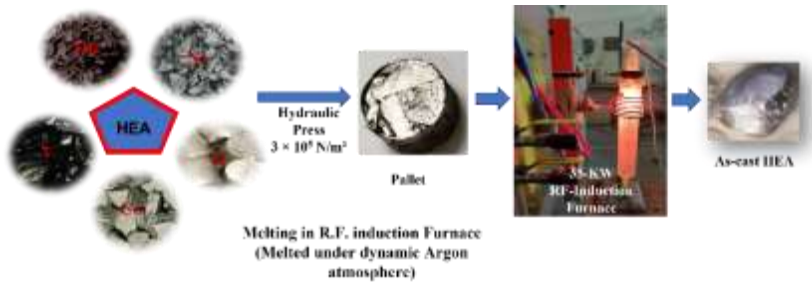

**Figure 1**: Presents a schematic diagram illustrating the synthesis protocol employed for producing the TiVCoNiMn$_2$ high-entropy alloy (HEA). This diagram visually represents the step-by-step procedure utilized in synthesizing the HEA mentioned above.

### 2.2 Results and Discussion:

Figure 2 shows the X-ray diffraction (XRD) patterns obtained from the as-cast TiVCoNiMn$_2$ high-entropy alloy (HEA). The diffraction profile was obtained to perform a comprehensive structural analysis of the as-cast and anneal HEAs sample. Figure 2 a shows the XRD profile of the as-cast HEA, figure 2 b shows the annealed HEA of 1000°C for 24 hours. The as-cast HEA exhibits a body-centered cubic (BCC) phase and the lattice parameter 2.95 Å. Following a thermal treatment involving annealing at a temperature of 1000 °C for 24 hours, the resultant material exhibits a body-centered cubic (BCC) crystal structure accompanied by some intermetallic phases. The GOF of the as-cast is 1.002, and for annealed is 0.98. These GOFs validate that all profiles fit perfectly.

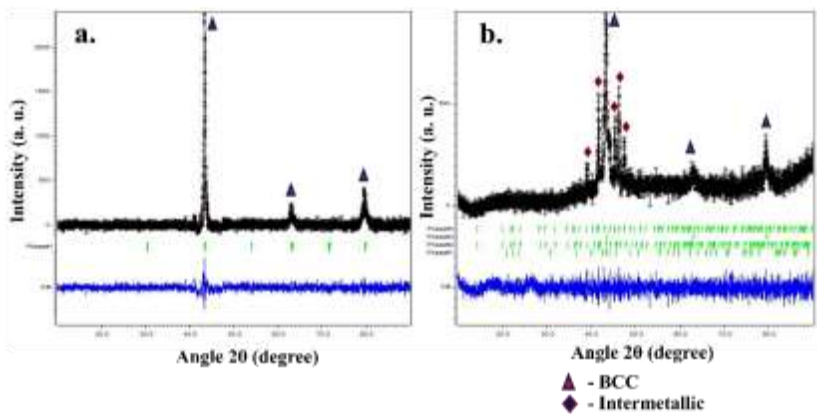

**Figure 2:** The X-ray diffraction (XRD) patterns of the as-cast TiVCoNiMn$_2$ high-entropy alloy (HEA) and the annealed TiVCoNiMn$_2$ HEA at 1000°C for 24 hours are presented in Figure a and Figure b, respectively.

**2.3 Optical images of polished surface:**

The alloy surfaces were subjected to cloth polishing, and subsequent optical imaging revealed the absence of any visible cracks. Figure a displays an optical image of the as-cast high-entropy alloy (HEA), while figure 2 b showcases an optical image of the heat-treated HEA. In figure 2 a, small black patches are observed, showing the presence of additional miner intermetallic phases alongside the body-centered cubic (BCC) structure. In Figure B, a distinct manifestation of the double phase is observed, wherein one phase is characterized by a body-centered cubic (BCC) structure. In contrast, the other phase exhibits an intermetallic phase. This multiphase configuration is prominently displayed along the grain boundary.

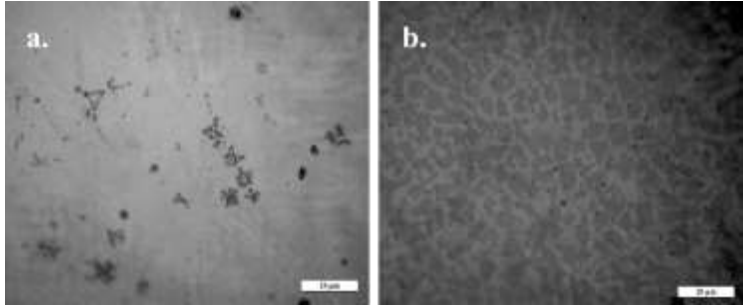

**Figure 2 a.** presents an optical image depicting the as-cast high-entropy alloy (HEA). Figure b displays an optical image depicting the microstructure of a heat-treated High Entropy Alloy (HEA) at 1000 ◦C for 24 hours.

**2.4 SEM Analysis:**

The present study presents the findings obtained from the utilization of Scanning Electron Microscopy (SEM) in conjunction with Energy Dispersive X-ray Analysis (EDAX) for the investigation of the microstructural and elemental properties of the TiVCoNiMn$_2$ high entropy alloy. These results provide significant insights into the alloy's composition and elemental analysis. Scanning electron microscopy (SEM) micrographs comprehensively depict the alloy's surface morphology, revealing a uniform and finely distributed microstructure. The elemental composition of the alloy was further verified through EDX analysis, providing additional validation. In figure 3 shows the SEM image with corresponding ESX spectra. The analysis confirmed the presence of titanium (Ti), vanadium (V), manganese (Mn), cobalt (Co), and nickel (Ni) in approximate atomic percentages of 16.67%, 16.67%, 33.32%, 16.67%, and 16.67% is after melting is 1.90%, 14.7%, 33.5%, 18.4%, 16.4% respectively sown in Table 1. The results provide significant insights into the properties of the alloy mentioned above. These micrographs reveal a consistent and finely dispersed microstructure across the alloy's surface.

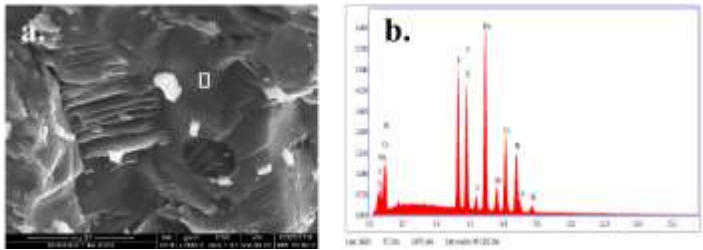

**Figure 3** shows a scanning electron microscopy (SEM) micrograph. Figure b presents the elemental spectra. These spectra depict the distribution and abundance of various elements in the sample.

| Element | Ti | V | Mn | Co | Ni |
|---|---|---|---|---|---|
| Atomic % | 16.9 | 14.7 | 33.5 | 18.4 | 16.4 |

**Table 1:** Elemental atomic presence of constituent elements in the TiVCoNiMn$_2$ high entropy alloy.

**2.5 TEM Analysis:**

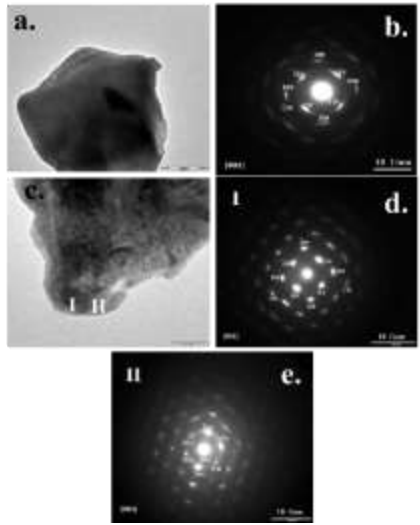

To corroborate the structural analysis of the X-ray diffraction (XRD) results, an additional characterization technique, namely transmission electron microscopy (TEM), was employed to investigate the phase composition and microstructure of the TiVCoNiMn$_2$ high-entropy alloy (HEA). The bright field transmission electron microscopy (TEM) micrograph of the as-synthesized and annealed high-entropy alloy (HEA), as depicted in Figure 4(a, c), reveals the absence of any additional phases in as-cast HEAs. Still, figure c shows different contrast that should be multiple phase formation on annealing. The structural analysis of the as-cast high-entropy alloy (HEA) depicted in figure 4(b, d, e) reveals the BCC phase with the presence of additional hexagonal structure base intermetallic phase of heat-treated HEA. The corresponding selected area diffraction (SAD) pattern further supports this classification, revealing a space group of P6$_3$/mmc for this HEA system.

**Figure 4:** (a, c) TEM bright field micrograph of as-cast and heat-treated HEA synthesized by RF induction melting (b, d, e) Corresponding SAD patterns are indexed with BCC and hexagonal structure type intermetallic.

### 2.6 Vickers Hardness Analysis:

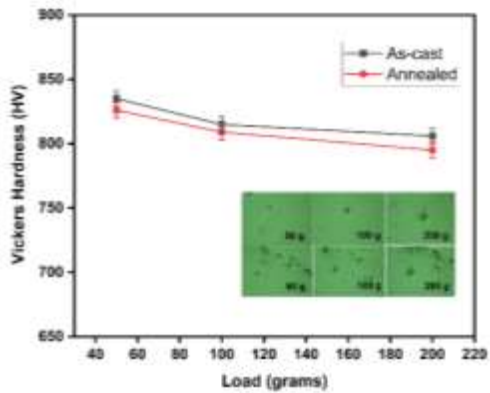

A load of 200 grams was carefully applied to test the Vickers hard of the present system. The dwell time, which refers to the duration for which the load was maintained, was precisely set at 15 seconds. The investigation aimed to assess heat treatment's impact on the hardness of as-cast and heat-treated High Entropy Alloys (HEAs). The results obtained from the experimental analysis revealed no significant alteration in the hardness of the HEAs between the as-cast and heat-treated conditions. In fugure 5 we can clearly see that hardness did not differ between annealed (795 HV) and as-cast (806 HV).

**Figure 4:** HV test by using Vickers hardness tester varying the load 50, 100, 200 grams with 15 second dwell time and inside the figure shows indentation image after indenting.

### 3. Conclusions:

The TiVCoNiMn$_2$ High Entropy Alloy (HEA) was synthesized. Based on the comprehensive analysis of X-ray diffraction (XRD), transmission electron microscopy (TEM), and optical imaging, it has been observed that the annealed high-entropy alloy (HEA) subjected to a temperature of 1000 °C for 24 hours has transformed into an intermetallic structure characterized by a BCC phase. Notably, this heat treatment did not induce any significant alteration in the hardness of the material. This reported HEA is derived from both hydride and non-hydride-producing elements. This could be a useful hydrogen storage material. The hydrogen absorption/desorption capabilities of these HEAs are promising.

## References

[1] T.P. Yadav, S. Mukhopadhyay, S.S. Mishra, N.K. Mukhopadhyay, O.N. Srivastava (2017) Synthesis of a single phase of high-entropy Laves intermetallics in the Ti–Zr–V–Cr–Ni equiatomic alloy. Phil. Mag. Lett.97 (12), 494-503.
[2] S.S. Mishra, S. Mukhopadhyay, T.P. Yadav, N.K. Mukhopadhyay, O.N. Srivastava (2019) Synthesis and characterization of hexanaryTi–Zr–V–Cr–Ni–Fe high-entropy Laves phase. J. Mater. Res.34 (5), 807-818.
[3] S.S. Mishra, T.P. Yadav, O.N. Srivastava, N.K. Mukhopadhyay, K. Biswas (2020) Formation and stability of C14 type Laves phase in multi component high-entropy alloys. J. Alloy. and Com.832, 153764

[4] T.P. Yadav, A. Kumar, M.A. Shaz, N.K. Mukhopadhyay (2022) High-Entropy Alloys for Solid Hydrogen Storage: Potentials and Prospects. Transactions of the Indian National Academy of Engineering, 7 147-156.

[5] A. Kumar, T.P. Yadav, M.A. Shaz, N.K. Mukhopadhyay (2023) Hydrogen Storage Performance of C14 Type $Ti_{0.24}V_{0.17}Zr_{0.17}Mn_{0.17}Co_{0.17}Fe_{0.08}$ High Entropy Intermetallics. Transactions of the Indian National Academy of Engineering.

[6] A. Kumar, T.P. Yadav, N.K. Mukhopadhyay (2022) Notable hydrogen storage in Ti–Zr–V–Cr–Ni high entropy alloy. International Journal of Hydrogen Energy, 47, 22893-22900.

[7] A. Kumar, T.P. Yadav, M.A. Shaz, N.K. Mukhopadhyay (2024) Hydrogen storage properties in rapidly solidified TiZrVCrNi high entropy alloys. Energy Storage.

[8] A. Kumar, T.P. Yadav, N.K. Mukhopadhyay (2024) Chapter 3.2, Hydrogen storage in high entropy alloys. Towards Hydrogen Infrastructure, 133-164.

[9] D.B. Miracle, O.N. Senkov (2017) A critical review of high entropy alloys and related concepts. ActaMaterialia122, 448-511.

[10] L.B. Chen, R. Wei, K. Tang, J. Zhang, F. Jiang, L. He, J. Sun (2018) Heavy carbon alloyed FCC-structured high entropy alloy with excellent combination of strength and ductility. Materials Science andEngineering: A 716, 150–156.

[11] L. Zhang, X. Huo, A. Wang, X. Du, L. Zhang, W. Li, N. Zou, G. Wan, G. Duan, B. Wu (2020) A ductile high entropy alloy strengthened by nano sigma phase. Intermetallics122, 106813.

[12] S. Lu, Y. Zu, G. Chen, B. Zhao, X. Fu, W. Zhou (2020) A multiple nonmetallic atoms co-doped CrMoNbWTi refractory highentropy alloy with ultra-high strength and hardness. Materials Science and Engineering: A 795, 140035.

[13] C. Zhu, Z. Li, C. Hong, P. Dai, J. Chen (2020) Microstructure and mechanical properties of the TiZrNbMoTa refractory highentropy alloy produced by mechanical alloying and spark plasma sintering. International Journal of Refractory Metals and Hard Materials 93, 105357.

[14] S. Guo, C. T. Liu (2011) Phase stability in high entropy alloys: formation of solid-solution phase or amorphous phase. Progress in Nat ural Science: Materials International 21(6), 433–46.

[15] S. Guo, C. Ng, J. Lu, C.T. Liu (2011) Effect of valence electron concentration on stability of fcc or bcc phase in high entropy alloys. Journal of Applied Physics 109(10), 103505–103542.

[16] C. Wang, C.L. Tracy , S. Park, J. Liu, F. Ke, F. Zhang, T. Yang, S. Xia, C. Li, Y. Wang, Y. Zhang, W.L. Mao, R.C. Ewing (2019) Phase transformations of Al-bearing high entropy alloys $Al_xCoCrFeNi$ (x = 0, 0.1, 0.3, 0.75, 1.5) at high pressure. Appl. Phys. Lett.114, 091902.

[17] Z. Guan, C. Feng, H. Song, Y. Zhang, F. Zhang (2022) Effects of Ge addition on the structure, mechanical and magnetic properties of $(CoCrFeNi)_{100-x}Ge_x$ high-entropy alloys. Phys. Scri.97, 125701.